\documentclass[%
 reprint,
 amsmath,amssymb,
 aps,
 prl,
]{revtex4-1}

\usepackage[dvips]{graphicx}
\usepackage{dcolumn}
\usepackage{bm}

\usepackage{color,graphicx}
\usepackage[colorlinks=true,urlcolor=blue,citecolor=blue,linkcolor=blue]{hyperref}
\usepackage{epstopdf} 
\usepackage{pdfpages}

\newcommand{\WV}[1]{\textcolor{black}{#1}}

\newcommand{\BKR}[1]{\textcolor{black}{#1}}
\newcommand{\BKD}[1]{\textcolor{black}{#1}}

\begin{document}
\preprint{APS/123-QED}

\title{Unbalanced Homodyne Correlation Measurements}

\author{B. K\"uhn}
 
\author{W. Vogel}

\affiliation{Arbeitsgruppe Quantenoptik, Institut f\"ur Physik, Universit\"at Rostock, D-18051 Rostock, Germany}

\date{\today}

\begin{abstract}

A method is introduced that allows one to measure normal-ordered moments of the displaced photon-number operator up to higher orders, without the need of photon-number resolving detectors. 
It is based on unbalanced homodyne correlation measurements, with the local oscillator being replaced by a displaced dephased laser. 
The measured moments yield a simple approximation of quasiprobabilities, representing
the full quantum state. Quantum properties of light are efficiently certified through normal-ordered observables 
directly accessible by our method, which is illustrated for a weakly squeezed vacuum and a single-photon-added thermal state.

\begin{description}
\item[PACS numbers]42.50.Dv, 03.65.Ta, 42.50.Lc, 03.65.Wj
\end{description}
\end{abstract}

\pacs{Valid PACS appear here}
\maketitle

\paragraph{Introduction.---}

Phase-sensitive measurements of light play a key role for any application of radiation fields, both in classical and quantum physics, or technology. 
For this purpose, homodyne measurement techniques were introduced~\cite{YS78}. They rely on the superposition of the signal with a coherent reference field---the local oscillator (LO). 
Excess noise of the usually strong LO can be eliminated via balanced homodyne detection~\cite{YC83,ACY83}. 
In quantum optics this opened the possibility to fully characterize quantum states of light~\cite{SBRF93,HaNP,WiNP,Bim1}; 
for reviews see Refs.~\cite{WVO99,L05,LR09}. 
Unbalanced homodyne detection with a weak LO yields a simple determination of the quantum state~\cite{Roy1,Wal1}. 
As it needs to distinguish adjacent photon numbers, it could be realized for extremely weak signals only~\cite{Ban1}.

Another important development was the application of correlation measurement techniques 
for the experimental demonstration of the  quantum nature of light through the antibunching of photons in atomic resonance fluorescence~\cite{KDM77}. 
A strong point is that the measured intensity correlations are normal and time ordered, and hence insensitive to vacuum-noise effects. 
For the measurement of phase-sensitive squeezing in atomic resonance fluorescence, small quantum efficiencies in homodyne detection pose a severe problem~\cite{Ma82}. 
To overcome this deficiency, homodyne correlation measurements with a weak LO were proposed~\cite{Vo91,Vo95}. 
Only very recently this technique has been applied to detect squeezing in resonance fluorescence from a semiconductor quantum dot~\cite{Sch1}; see also Refs.~\cite{Car1,Fos1,Ger1}
for related techniques. 
These methods were further developed to combine the advantages of homodyne correlation and balanced measurements~\cite{Shc2}. 
Using a strong LO, normal-ordered quadrature moments of high orders can be detected by this balanced homodyne correlation technique.

In \BKR{the present} Letter we introduce an unbalanced homodyne correlation measurement (UHCM) technique. 
\BKR{This method allows for a direct detection of normal-ordered moments of the displaced photon-number operator by linear standard detectors and, thus, it overcomes the limitations of photon-number resolving detectors. } 
\BKR{It unifies }the simplicity of quantum state characterization by unbalanced homodyne detection with the loss insensitivity of homodyne correlation measurements. 
\BKR{This is achieved by replacing the standard LO by a novel reference field, namely a coherently displaced dephased laser (DDL).} 
\BKR{Our method will prove beneficial to verify nonclassical phenomena of light by properly constructed and directly accessible nonclassicality witnesses.}

\paragraph{Measurement technique.---}
The state of single mode light can be fully represented by the $s$-parametrized quasiprobabilities of Cahill and Glauber~\cite{Cah1}, which can be given in the simple form~\cite{Wal1}
\begin{eqnarray}\label{eq:sP}
P(\alpha;s)=\dfrac{2}{\pi(1-s)}\sum_{\BKR{n}=0}^\infty\left[\dfrac{\eta(1-s)-2}{\eta(1-s)}\right]^{\BKR{n}}p_{\BKR{n}}(\alpha),
\end{eqnarray}
with $s$ being the ordering parameter and $\eta$ the quantum efficiency.
It depends on the photoelectric counting statistics $p_{\BKR{n}}(\alpha)$ of the coherently displaced quantum state, $\hat\rho \mapsto \hat\rho(-\alpha)$, 
\begin{eqnarray}\label{eq:pk}
p_{\BKR{n}}(\alpha)&=&\left\langle:\dfrac{(\eta\hat n(\alpha))^{\BKR{n}}}{\BKR{n}!}e^{-\eta\hat n(\alpha)}:\right\rangle.
\end{eqnarray}
Here, $\hat n(\alpha)=(\hat a^\dagger-\alpha^*)(\hat a-\alpha)$ is the displaced photon-number operator. 
The symbol $:\cdot:$ denotes normal ordering of the annihilation and creation operators $\hat a$  and $\hat a^\dagger$, respectively.
Even in the case of imperfect detection, for quantum efficiencies $\eta<1$, the quantum state is obtained through Eq.~\eqref{eq:sP} from the measured statistics.

The unbalanced homodyne detection method for the experimental state reconstruction is based on the measurement of the displaced photon-number statistics~\eqref{eq:pk}. 
Approaches were introduced in Ref.~\cite{Roy1} for the Wigner function, 
in Ref.~\cite{Wal1} for general $s$-parametrized quasiprobabilities, and in Ref.~\cite{Kie2} for more general regular quasiprobabilities. In principle, the experimental implementation of 
these state reconstructions requires photon-number resolving detectors. 
\BKR{Even novel techniques, such as superconducting nanowire~\cite{Ale1,Jah1} or transition-edge~\cite{Sm1,Cal1} detectors, only resolve moderate photon numbers, so that a truncation of the photon number in Eq.~\eqref{eq:sP} is indispensable. This necessitates 
involved studies of systematic errors~\cite{Kie2}. 
\BKD{In this Letter we solve the problem by focussing on normal-ordered moments of the displaced photon number. 
The knowledge of all these moments determines the statistics~\eqref{eq:pk} via
\begin{eqnarray}
p_{n}(\alpha)&=&\sum_{\ell=n}^\infty(-1)^{\ell-n}\dfrac{\eta^{\ell}}{(\ell-n)!}\dfrac1{n!}\langle:[\hat n(\alpha)]^\ell:\rangle,
\end{eqnarray} 
from which the full state representation in terms of quasiprobabilities~\eqref{eq:sP} is derived.
As we will see below, a truncation of the quasiprobabilities, in \WV{the order of moments $\langle:[\hat n(\alpha)]^m:\rangle$ rather than in the} photon number, does not introduce any systematic error with respect to the certification of nonclassical effects. 
Moreover, our UHCM technique directly yields the moments $\langle:[\hat n(\alpha)]^m:\rangle$ 
with linear standard detectors which even applies to large photon numbers.}} 

\begin{figure}[h]
\hbox{\hspace{0.3cm}\includegraphics[clip,scale=0.58]{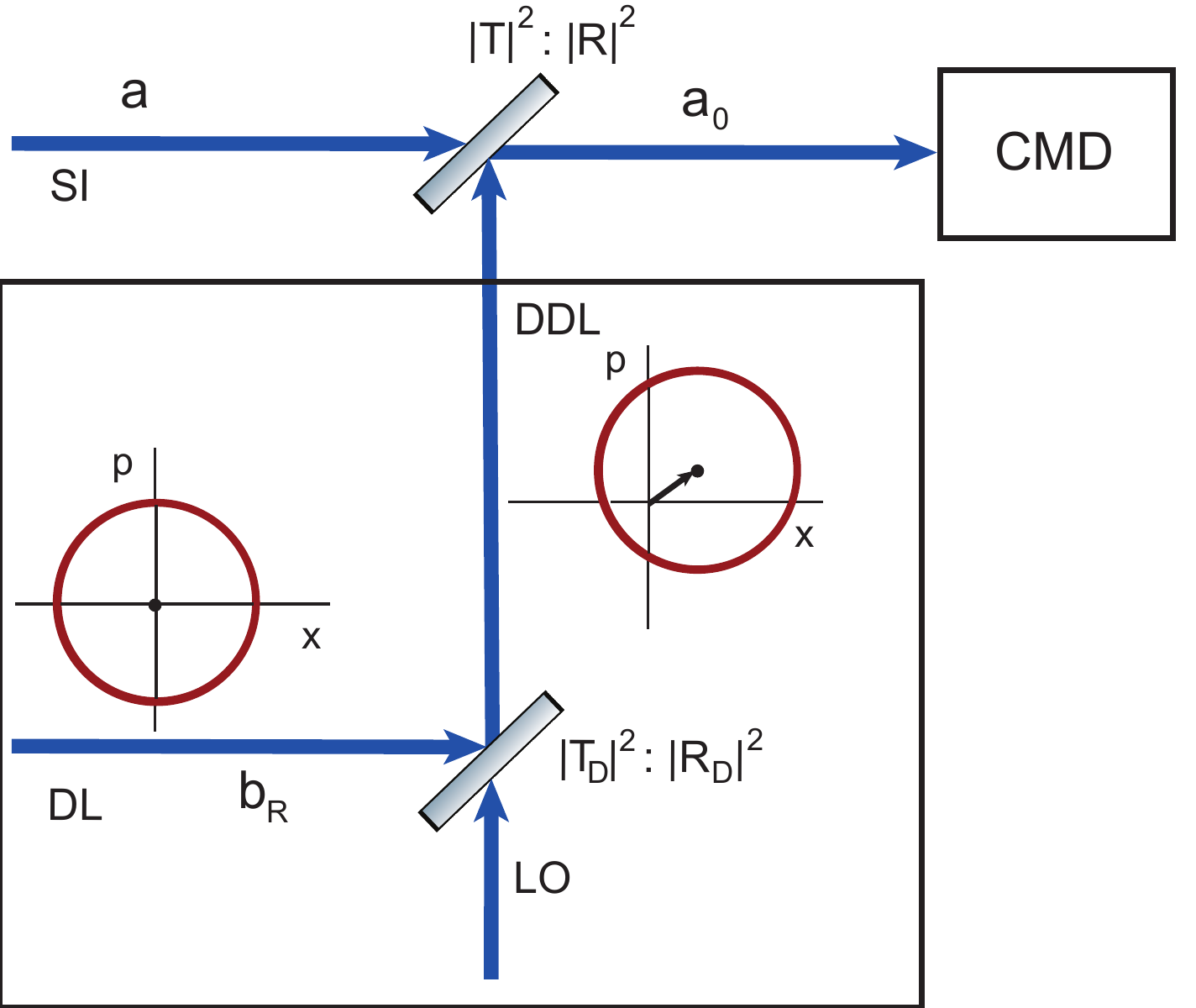}}
\caption{Experimental UHCM setup. The signal (SI) is combined with a displaced dephased laser (DDL) at a highly transmitting beam splitter. 
The output $\hat a_0$ is recorded via a correlation measurement device (CMD), which yields the moments $\langle:[\hat n(\alpha)]^m:\rangle$.}
\label{fig:Setup1}
\end{figure}

The basic setup of our measurement technique is illustrated in Fig.~\ref{fig:Setup1}. 
The signal $\hat a$ is combined with the displaced dephased laser (DDL) by a beam splitter of high transmissivity, with amplitude transmittance and reflectance $T$ and $R$, respectively.  
The DDL replaces the local oscillator used in standard homodyne detection. 
This unusual reference field is used to accomplish that the moments $\langle:[\hat n(\alpha)]^m:\rangle$ are directly extracted from the detection 
of the output field $\hat a_0$ of the beam splitter with a correlation measurement device (CMD).
The DDL is produced, by combining a mode $\hat b_{\rm R}$, prepared in a strong and fully dephased coherent state, 
$\int_0^{2\pi} d\phi\,|\beta_{\rm R}e^{i\phi}\rangle\langle\beta_{\rm R}e^{i\phi}|/2\pi$, $\beta_{\rm R}\in\mathbb{R}$, with 
a comparatively weak local oscillator of complex amplitude $\beta_{\rm D}$ at a second beam splitter with amplitude transmittance $T_{\rm D}$ and reflectance $R_{\rm D}$. 
In phase space, the DDL has the form of a circle with the radius $|R_{\rm D}|\beta_{\rm R}$, which is much greater than the displacement amplitude, $|T_{\rm D}\beta_{\rm D}|$, of the circle center.
The DDL eliminates the coherent terms in the quadratures detected in balanced homodyne correlation measurements; see Ref.~\cite{Shc2}. 
Its strong incoherent amplitude, $\beta_{\rm R}$, guarantees a sufficiently strong field at the CMD to record moments up to higher orders.
The signal displacement caused by the DDL, $\alpha=-RT_{\rm D}\beta_{\rm D}/T$, controls the position in phase space, as desired in unbalanced homodyne detection. 
The signal $\hat a$ undergoes the transformation
\begin{eqnarray}\label{eq:disp}
\hat a_0=T\left(\hat a-\alpha\right)+RR_{\rm D}\hat b_{\rm R},
\end{eqnarray}
which yields the field in the input channel of the CMD.

\begin{figure}[h]
\hbox{\hspace{0.4cm}\includegraphics[clip,scale=0.6]{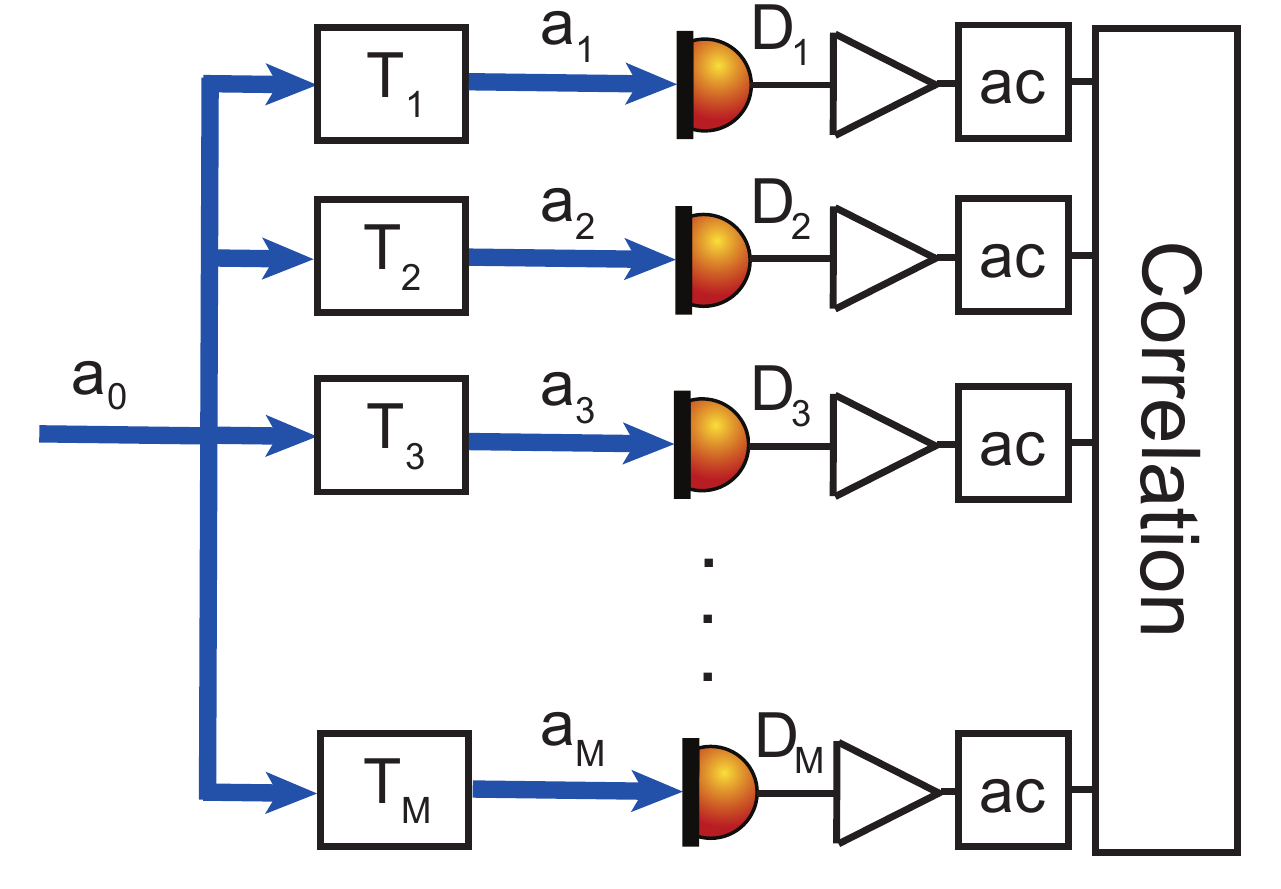}}
\caption{Structure of the CMD. The input mode is divided into $M$ modes, each one being recorded by a detector. 
The amplitude in each channel is attenuated by an individual factor $T_u$, $u=1,\dots, M$. The correlations in the AC parts of the amplified detector currents of different detectors are evaluated. 
}
\label{fig:Setup2}
\end{figure}

The structure of the CMD is shown in Fig.~\ref{fig:Setup2}.
The mode $\hat a_0$ is split into $M$ output modes $\hat a_u$, $u=1,\dots,M$. 
If $T_u$ denotes the overall amplitude transmission 
through the path from mode $\hat a_0$ to mode $\hat a_u$, the simple relation
\begin{eqnarray}\label{eq:plusvac}
\hat a_u=T_u\hat a_0+\mathrm{vac}
\end{eqnarray}
holds true.
The intensities of these modes are recorded by linear photodetectors 
with quantum efficiencies $\eta_u$. 
The electronic output signal of each detector is linearly amplified by a factor $g_u$, resulting
in an electric current $c_u$.
For our method to work properly, it is required that the mean currents, $\overline{c}_u$, are equal for all detectors, i.e.,
$g_u\eta_u|T_u|^2=\zeta$, where $\zeta$ is an arbitrary but fixed positive constant. 
This is easily achieved by adjusting the individual gain factors $g_u$ of the classical output currents of the detectors.

The measurement technique uses only the detected alternating currents (ac), labeled as $\tilde c_u$.
\BKR{Based on $N$ ac-data points, $\{\tilde c_{u,j}\}_{j=1}^N$, from each detector $D_u$, the equal-time ac-correlation of 
$2m$ detectors is
\begin{eqnarray}
\overline{\tilde c_1\cdots \tilde c_{2m}}=\dfrac1{N}\sum_{j=1}^N \tilde c_{1,j}\cdots\tilde c_{2m,j}.
\end{eqnarray}
According to Sec.~III of the Supplemental Material, this quantity is related to the normal-ordered displaced number moment of power $m$,
\begin{equation}\label{eq:mom}
\Gamma_{m,\alpha}=\overline{\tilde c_1\cdots \tilde c_{2m}}=\binom{2m}{m}\tilde\zeta^{2m}\left\langle:\left[\hat n(\alpha)\right]^m:\right\rangle,
\end{equation}
with $\tilde\zeta=\zeta|T||R||R_{\rm D}|\beta_{\rm R}$.} 
\BKR{Herein the detector efficiencies $\eta_u$ appear only as prefactors, as vacuum effects are not detected in the correlation measurement.} Equation~\eqref{eq:mom} is a central result of this Letter. The effect of amplitude fluctuations of the DDL are studied in Sec.~VII of the Supplemental Material.
\BKR{The dephasing of the DDL can be realized during the data sampling, by continuous phase shifting. As long as the phase distribution is uniform, 
possible short-time coherences do not affect the measurement outcome. This can be realised by a phase postprocessing through quantum random number generation based on the detection 
of vacuum noise, as performed in~\cite{Agu1}.}

A strength of our technique compared to balanced homodyne detection is that no reference measurements like the vacuum detection for proper normalization is required, which would give rise to statistical uncertainties.
Remarkably, even the dark noise appearing in each detector independently is erased by the UHCM technique. For a proof of this statement we refer to Sec.~IV of the Supplemental Material. 
\BKD{However, \WV{occasionally occurring correlated detector dark counts, e.g., due to} background light or current fluctuations in a common power supply, require a more careful analysis.}  
Most importantly, our method does not require photon-number resolving detectors. 

It is noteworthy that the generalization of our method to multimode radiation and even to space-time dependent correlation measurements is straightforward. 
In principle, one may use for each radiation mode a CMD of the type under study. 
The determination of space-time dependent quantum correlations of light~\cite{Vo08} needs  control of the relative time delays in the recorded data. 

\paragraph{Certification of nonclassicality.---}
\label{ch:nonclassicalitywitness}
A possible application of the UHCM technique is the certification of nonclassicality.
For this purpose we consider observables of the form $\hat W=\,:\hat f^\dagger(\hat a,\hat a^\dagger)\hat f(\hat a,\hat a^\dagger):$.
Whenever its expectation value is negative, the state cannot be represented as a mixture of coherent states,
$\hat\rho=\int d^2\alpha\,P_{\rm cl}(\alpha)|\alpha\rangle\langle\alpha|$, with a classical
probability distribution $P_{\rm cl}(\alpha)$. Such states are referred to as nonclassical ones~\cite{Tit1,Man1}.

Let us consider the operator
\begin{eqnarray}
\hat f^{(k)}_{w}(\alpha)=\sum_{m=0}^kh^{\BKR{(k)}}_mw^{2m}\left[\hat n(\alpha)\right]^m, 
\end{eqnarray}
with a positive constant $w$ and the \BKR{$(k+1)$-dimensional} unit vector $\boldsymbol{h}^{\BKR{(k)}}=(h_0,\dots,h_k)^T$, weighting the powers of the displaced photon-number operator \BKR{up to order $k$}.
Our UHCM technique detects normal-ordered moments of the displaced photon number. 
Thus, the function 
\begin{eqnarray}\label{eq:Fwk}
F^{(k)}_{w}(\alpha)=\langle:\hat f^{(k)\dagger}_{w}(\alpha)\hat f^{(k)}_{w}(\alpha):\rangle=\boldsymbol{h}^{\BKR{(k)}T}\underline{\boldsymbol{L}}^{\BKR{(k)}}_{\BKR{w}}(\alpha)\boldsymbol{h}^{\BKR{(k)}}
\end{eqnarray}
can be determined by the symmetric matrix of moments,
\begin{eqnarray}\label{eq:momentmatrix}
[\underline{\boldsymbol{L}}^{\BKR{(k)}}_{\BKR{w}}(\alpha)]_{mm'}=w^{2(m+m')}\langle:\left[\hat n(\alpha)\right]^{m+m'}:\rangle, 
\end{eqnarray}
with $m,m'=0,\dots,k$.
The function~\eqref{eq:Fwk} requires the moments $\langle:[\hat n(\alpha)]^z:\rangle$ for $z=1,\dots,2k$, which are accessible with $4k$ detectors.
\BKR{The matrix elements of $\boldsymbol{\underline{L}}^{\BKR{(k)}}_{\BKR{w}}(\alpha)$ follow from the measured correlations, $\Gamma_{m,\alpha}$ in Eq.~\eqref{eq:mom}, as
\begin{eqnarray}\label{eq:exptotheo}
[\boldsymbol{\underline{L}}^{\BKR{(k)}}_{\BKR{w}}(\alpha)]_{mm'}=\dfrac{\tilde w^{2(m+m')}}{\binom{2(m+m')}{m+m'}}\Gamma_{m+m',\alpha}
\end{eqnarray}
with the rescaled parameter $\tilde w=w/\tilde\zeta$.} 
If $F^{(k)}_{w}(\alpha)<0$ for some $\alpha$, the state is nonclassical.

Interestingly, the function $F^{\BKR{(\infty)}}_{w}(\alpha)$ represents the regularized quasiprobabilities $P_w(\alpha)$ in phase space 
if one replaces in Eq.~\eqref{eq:Fwk} $\boldsymbol{h}^{\BKR{(\infty)}} \mapsto \boldsymbol{H}$, 
with the components
\begin{eqnarray}\label{eq:quasiprobh}
H_{m}&=&\dfrac{2^{1/q+1/2}w}{\sqrt{\pi q\Gamma\left(2/q\right)}}\dfrac{(-1)^m}{(m!)^2}\Gamma\left(\dfrac{2}{q}(m+1)\right),
\end{eqnarray}
with $m=0,\dots,\infty$, $\Gamma(\cdot)$ the gamma function, and $q\in(2,\infty)$; for details see Sec.~II of the Supplemental Material.  
The function $P_w(\alpha)$ \BKR{is not only a full state representation of the signal field}, it also provides a universal nonclassicality test~\cite{Kie3,Kue1}.
It approaches the Glauber-Sudarshan $P$~function~\cite{Gla1,Sud1} in the limit $w\to\infty$; see Ref.~\cite{Kie3} and Sec.~I of the Supplemental Material.
The regularized $P$~function, $P_w$, was successfully reconstructed via balanced homodyne detection~\cite{Kie4,Kie5}. 
Nonclassicality is certified by negativities of this quasiprobability. 
However, such a reconstruction is sophisticated and requires the numerical calculation of pattern functions, cf. the Supplemental Material of Ref.~\cite{Kie4}.

Partial reconstruction of the regularized $P$~function is feasible by applying in Eq.~\eqref{eq:Fwk} for $\boldsymbol{h}^{\BKR{(k)}}$ the ($k+1$)-dimensional unit vector $\boldsymbol{\widetilde H}^{(k)}=\boldsymbol{H}^{(k)}/\Vert\boldsymbol{H}^{(k)}\Vert$, 
where the components of $\boldsymbol{H}^{(k)}$ are equal to the first $(k+1)$ components of $\boldsymbol{H}$, cf. Eq.~(\ref{eq:quasiprobh}). The quantity
\begin{eqnarray}\label{eq:tp}
\mathcal{P}^{(k)}_{w}(\alpha)=\boldsymbol{\widetilde H}^{(k)T}\underline{\boldsymbol{L}}^{\BKR{(k)}}_{\BKR{w}}(\alpha)\boldsymbol{\widetilde H}^{(k)}
\end{eqnarray}
is referred to as the truncated regularized $P$~function and is directly accessible via UHCM. 
\BKD{\WV{It is important that, in the case of our truncation procedure, the} negativities of $\mathcal{P}^{(k)}_{w}$ still certify nonclassicality.}
In principle, an infinite number of detectors yields a complete state reconstruction in terms of the regularized $P$~function. 

In general, the vector $\boldsymbol{\widetilde H}^{(k)}$ does not minimize $F^{(k)}_{w}(\alpha)$
in Eq.~\eqref{eq:Fwk}. 
To get an optimal nonclassicality test, we determine
the minimal eigenvalue of the matrix~\eqref{eq:momentmatrix},
\begin{eqnarray}\label{eq:minev}
\mathcal{F}^{(k)}_{w}(\alpha)=\min_{\boldsymbol{h}^{\BKR{(k)}}}F^{(k)}_{w}(\alpha)=\min\left\{\mathrm{S}[\underline{\boldsymbol{L}}^{\BKR{(k)}}_{\BKR{w}}(\alpha)]\right\}, 
\end{eqnarray}
where $\mathrm{S}[\cdot]$ denotes the spectrum of a matrix. Hence, it holds that $\mathcal{F}^{(k)}_{w}(\alpha)\leq\mathcal{P}^{(k)}_{w}(\alpha)$ and also $\mathcal{F}^{(k')}_{w}(\alpha)\leq\mathcal{F}^{(k)}_{w}(\alpha)$ for $k'>k$.
Recently, a similar approach was proposed for multiple on-off detectors~\cite{Spe1}.
The error calculation for $\mathcal{F}^{(k)}_{w}(\alpha)$ in Eq.~\eqref{eq:minev} is somewhat involved, cf. Sec.~V of the Supplemental Material. 

The above considerations are closely related to the matrix of moments approach of Agarwal and Tara~\cite{Aga1}. Its generalization to displaced photon numbers,
$[\,\boldsymbol{\underline{M}\,}(\alpha)]_{mm'}=\langle:\left[\hat n(\alpha)\right]^{m+m'}\!\!:\rangle$,
yields, in the limit of infinite matrix dimension, the full information about nonclassicality.
\BKR{The appearance of negativities of $\mathcal{P}^{(k)}_{w}(\alpha)$ depends on the value of $w$. 
The occurrence of negativities of $\mathcal{F}^{(k)}_{w}(\alpha)$, however, is independent of $w$.
The value of $w$ can be used to optimize the statistical significance.
Imperfect detection ($\eta_u<1$) can be compensated by adjusting $w$, cf.~Eq.~\eqref{eq:exptotheo}, since detector efficiencies are just prefactors in~Eq.~\eqref{eq:mom}.
Hence, precise knowledge of detector efficiencies is superfluous for certifying nonclassicality.}

\paragraph{Results for some nonclassical states.---}
The method described above is now applied to two different nonclassical states. 
The first one is a weakly squeezed vacuum state, $|\xi\rangle=\exp[(\xi^*\hat a^2-\xi\hat a^{\dagger 2})/2]|\mathrm{vac}\rangle$, with a relatively small value of $|\xi|=0.03$. 
Therefore, it is very close to the vacuum state. The Wigner function of a squeezed vacuum state is known to be non-negative and, thus, it does not certify nonclassicality by negative values.

\begin{figure}[h]
\hbox{\hspace{0.0cm}\includegraphics[clip,scale=0.48]{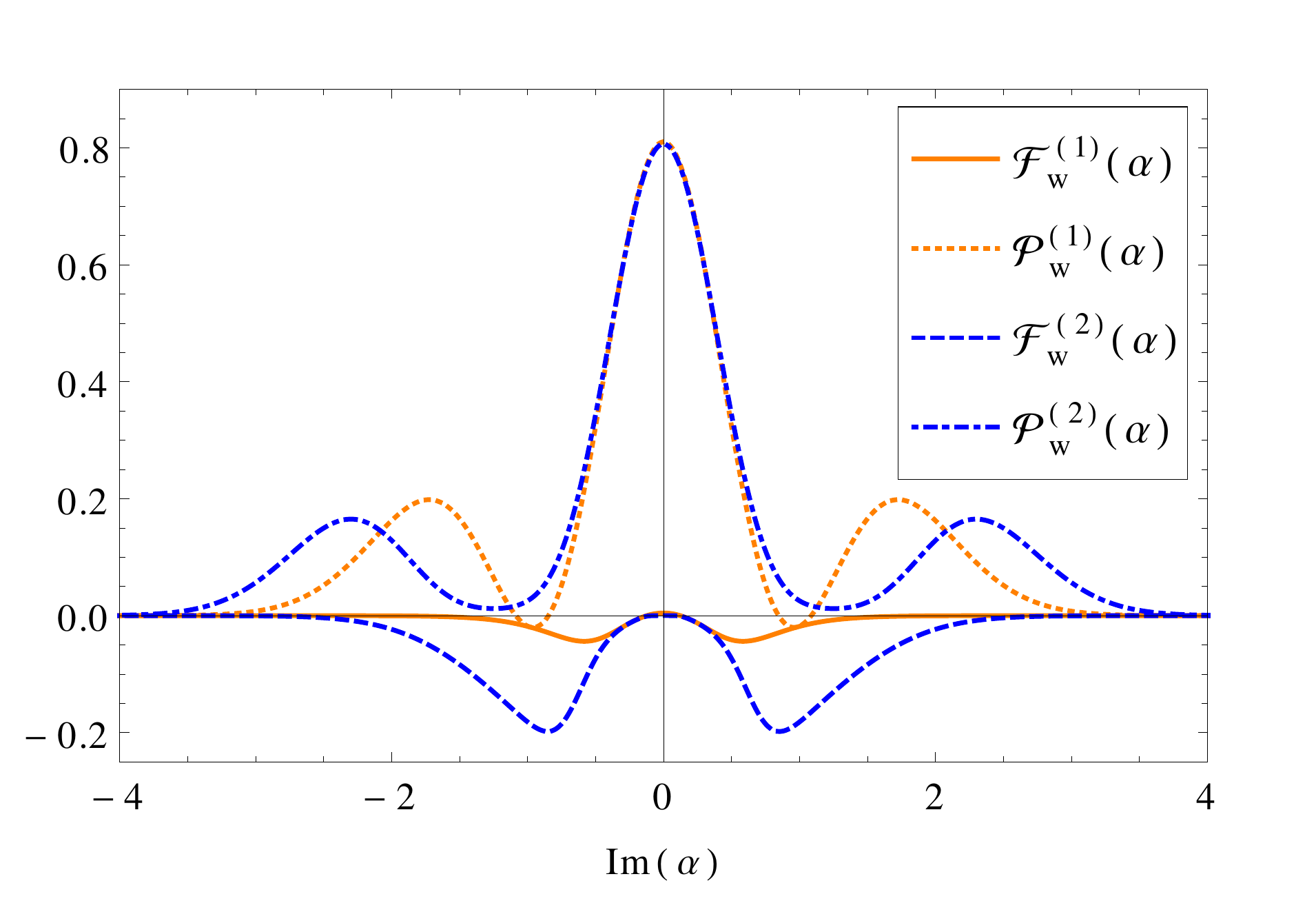}}
\caption{The minimal eigenvalues $\mathcal{F}^{(k)}_{w}(\alpha)$ and the truncated regularized $P$~function, $\mathcal{P}^{(k)}_{w}(\alpha)$, are shown \BKR{for $k=1,2$, $q=10$, and $w=1.5$;} for
a squeezed vacuum state with $\xi=0.03$. 
All functions are multiplied with $\exp[-|\alpha|^2]$. 
\BKR{Negative values certify nonclassicality.}}
\label{fig:sq}
\end{figure}

Figure~\ref{fig:sq} shows the truncated regularized \BKR{$P$~function~\eqref{eq:tp}} for $k=1,2$\BKR{, $q=10$, cf.~Eq.~\eqref{eq:quasiprobh}, and $w=1.5$} along the squeezing axis, as well as the minimal eigenvalues $\mathcal{F}^{(k)}_{w}(\alpha)$; see Eq.~\eqref{eq:minev}. 
\BKR{Negative values of these functions certify nonclassicality.}
Both functions are multiplied with the same Gaussian function to suppress their polynomial divergence for $|\alpha|\to \infty$.
The function $\mathcal{F}^{(1)}_{w}(\alpha)$ is negative in a much larger $\alpha$ interval than $\mathcal{P}^{(1)}_{w}(\alpha)$. For $k=2$, the truncated regularized $P$~function is
non-negative, whereas the minimal eigenvalue is clearly negative.
As $k$ is increased, $\mathcal{P}^{(k)}_{w}(\alpha)$ approaches the regularized $P$~function.
Remarkably, already $\mathcal{P}^{(1)}_{w}(\alpha)$ proves the nonclassicality very well. 
The negativities of $\mathcal{P}^{(1)}_{w}$ and $\mathcal{F}^{(1)}_{w}$ are for the same amount of data even more significant than those of $\mathcal{F}^{(2)}_{w}$; for details see Sec.~VI of the Supplemental Material.
This surprising result means that fewer detectors can even yield more insight than many detectors, also compare similar conclusions for measurements with multiple on-off detectors~\cite{Lui1}. \WV{As a general rule, to restrict the needed resources, a truncation is advisable whenever a significant certification of quantum effects has been achieved.}

\begin{figure}[h]
\vspace{-0.5cm}
\hbox{\hspace{0.0cm}\includegraphics[clip,scale=0.50]{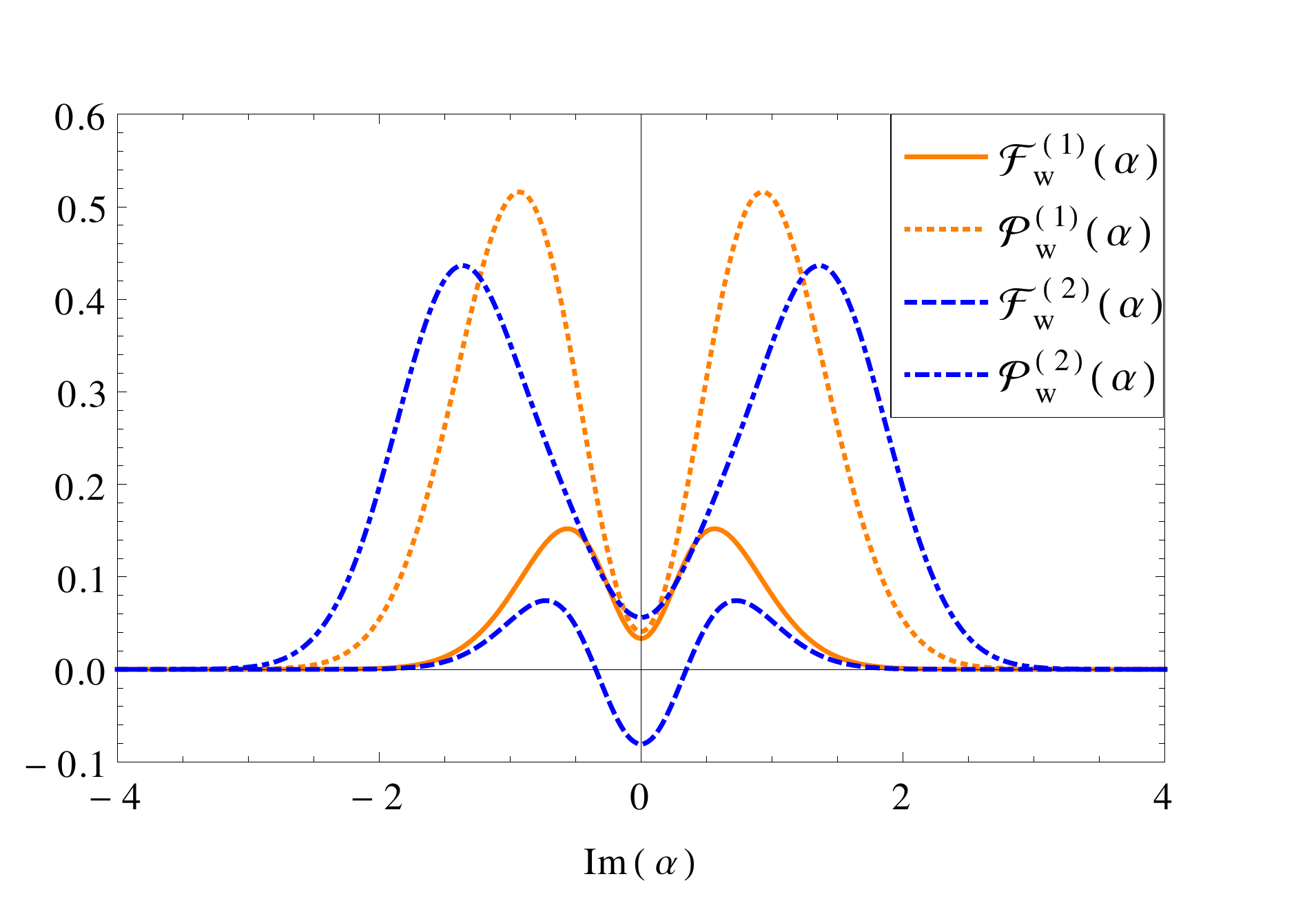}}
\caption{The minimal eigenvalues $\mathcal{F}^{(k)}_{w}(\alpha)$ and the truncated regularized $P$~function, $\mathcal{P}^{(k)}_{w}(\alpha)$, are shown \BKR{for $k=1,2$, $q=10$, and $w=1.3$;} for 
the SPATS with the parameters described in the text. 
All functions are multiplied with $\exp[-1.4|\alpha|^2]$. 
\BKR{Negative values certify nonclassicality.}
}
\label{fig:spats}
\end{figure}

As a second example, a single-photon-added thermal state (SPATS), $\hat\rho\propto \hat a^\dagger(\overline{n}/(1+\overline{n}))^{\hat n}\hat a$,  is considered with 
the mean photon number $\overline{n}$ of the thermal background. This state is comparable to a single photon state, which embodies the particle nature of light. 
For the overall quantum efficiency of $0.5$ and $\overline{n}=0.8$, several
nonclassicality conditions were tested~\cite{Bel1}, which 
did not certify quantumness. 
In Fig.~\ref{fig:spats} it is seen that the 
functions $\mathcal{P}^{(k)}_{w}$ are nonnegative for $w=1.3$\BKR{, $q=10$,} and $k=1,2$. 
However, the minimal eigenvalue $\mathcal{F}^{(k)}_{w}$ attains negative values for $k\geq 2$, uncovering the nonclassicality of this state.

\paragraph{Conclusions.---}
In conclusion, we have proposed unbalanced homodyne correlation measurements for a direct experimental determination of normal-ordered displaced photon-number moments. This becomes feasible by combining the signal field with a displaced dephased laser, then
recording the resulting light by a number of linear standard detectors, and correlating the alternating photocurrents. Based on these moments, one obtains
an approximate quasiprobability representation of the quantum state.
For our method, photon-number resolving detectors are not required.
The determination of the minimal eigenvalue of the matrix of normal-ordered displaced photon-number moments yields a powerful nonclassicality test,
as it is demonstrated for two examples. 

\paragraph*{Acknowledgements.--}



\begin{widetext}
\newpage


\section*{Supplementary material}

# Supplemental Material
# Unbalanced Homodyne Correlation Measurements

B. Kühn and W. Vogel

*Arbeitsgruppe Quantenoptik, Institut für Physik, Universität Rostock, D-18051 Rostock, Germany*

(Dated: April 20, 2016)

Section I gives a brief introduction to regularized $P$ functions and nonclassicality filters. Section II provides a witness operator which corresponds to the regularised $P$ function. Section III relates the equal-time correlation of alternating electric currents to normal-ordered displaced number moments. Section IV shows that dark noise, appearing independently in each of the detectors, is irrelevant in the considered unbalanced homodyne correlation measurement scheme. In the Section V, the error calculation is performed and statistical significances are considered in Section VI. In Sec. VII, amplitude fluctuations of the DDL are studied.

## I. REGULARIZED $P$ FUNCTION

It is known, that the Glauber-Sudarshan $P$ function is given by the Fourier transform of its characteristic function $\Phi(\beta)$, i.e.,

$$P(\alpha) = \frac{1}{\pi^2} \int d^2\beta \, \Phi(\beta) \, e^{\alpha\beta^* - \alpha^*\beta}. \tag{1}$$

Due to the fact, that the characteristic function $\Phi(\beta)$ for some states diverges for $|\beta| \to \infty$, the $P$ function is in general highly singular, which prevents its experimental reconstruction. This problem is completely solved by so-called nonclassicality filters, $\Omega_w(\beta)$. These functions map $\Phi(\beta)$ to a new characteristic function $\Omega_w(\beta)\Phi(\beta)$, which corresponds to a regularized $P$ function [1],

$$P_w(\alpha) = \frac{1}{\pi^2} \int d^2\beta \, \Omega_w(\beta) \, \Phi(\beta) \, e^{\alpha\beta^* - \alpha^*\beta}. \tag{2}$$

This phase-space distribution is a measurable counterpart of the $P$ function and is also referred to as nonclassicality quasiprobability. The nonnegativity of the Fourier transform of the filter guarantees that negativities of $P_w$ have its origin solely in the nonclassical character of the state and not in the filtering procedure. The regularized $P$ function approaches the $P$ function by increasing the filter width parameter, $w \to \infty$. Accordingly, a state is nonclassical, if there exists an $w < \infty$ and an $\alpha$, so that $P_w(\alpha) < 0$. Our approach applies autocorrelation filters of the form

$$\Omega_w(\beta) = \frac{q2^{2/q-1}}{\pi\Gamma(2/q)w^2} \int d^2\beta' \, e^{-(|\beta'|/w)^q - (|\beta+\beta'|/w)^q}. \tag{3}$$

The parameter $q \in (2, \infty)$ fixes a particular filter function and $\Gamma(\cdot)$ is the gamma function.

## II. WITNESSES CORRESPONDING TO REGULARIZED $P$ FUNCTIONS

The regularized $P$ function can be written as a convolution

$$P_w(\alpha) = \int d^2\gamma P(\gamma) \, \mathcal{F}[\Omega_w](\alpha - \gamma, \alpha^* - \gamma^*) \tag{4}$$

of the $P$ function with the Fourier transform $\mathcal{F}[\cdot]$ of the nonclassicality filter,

$$\Omega_w(\beta) = \int d^2\beta' \, \tilde{f}_w^*(\beta', \beta'^*) \, \tilde{f}_w(\beta + \beta', \beta^* + \beta'^*). \tag{5}$$

The latter is an autocorrelation function of the function

$$\tilde{f}_w(\beta, \beta^*) = \frac{1}{w}\sqrt{\frac{q \, 2^{2/q-1}}{\pi\Gamma(2/q)}} e^{-(|\beta|/w)^q}, \tag{6}$$

and, thus, its Fourier transform is given by

$$\mathcal{F}\left[\Omega_w\right](\alpha, \alpha^*) = \left|\mathcal{F}[\tilde{f}_w](\alpha, \alpha^*)\right|^2 = \langle \alpha | : \hat{f}_w^\dagger \hat{f}_w : |\alpha\rangle \qquad (7)$$

with the definition

$$\hat{f}_w = \mathcal{F}[\tilde{f}_w](\hat{a}, \hat{a}^\dagger). \qquad (8)$$

Applying relation (7), the regularized $P$ function (4) can be rewritten as

$$\begin{aligned} P_w(\alpha) &= \int d^2\gamma \, P(\gamma) \langle \gamma | \hat{D}(\alpha) : \hat{f}_w^\dagger \hat{f}_w : \hat{D}^\dagger(\alpha) | \gamma \rangle \\ &= \langle \hat{W}(\alpha) \rangle, \end{aligned} \qquad (9)$$

i.e., as the expectation value of a witness operator

$$\hat{W}(\alpha) = \hat{D}(\alpha) : \hat{f}_w^\dagger \hat{f}_w : \hat{D}^\dagger(\alpha). \qquad (10)$$

In order to determine the operator function $\hat{f}_w$ it is necessary to calculate the Fourier transform of (6). One obtains for this quantity after a straightforward calculation

$$\mathcal{F}[\tilde{f}_w](\alpha, \alpha^*) = w \frac{2^{1/q+1/2}}{\sqrt{\pi q \Gamma(2/q)}} \sum_{m=0}^{\infty} \frac{(-w^2 |\alpha|^2)^m}{(m!)^2} \Gamma\left(\frac{2}{q}(m+1)\right), \qquad (11)$$

and according to Eq. (8)

$$\hat{f}_w = w \frac{2^{1/q+1/2}}{\sqrt{\pi q \Gamma(2/q)}} \sum_{m=0}^{\infty} \frac{(-1)^m}{(m!)^2} \Gamma\left(\frac{2}{q}(m+1)\right) w^{2m} \hat{n}^m. \qquad (12)$$

With this last equation the witness operator (10) is completely determined.

### III. EQUAL-TIME CORRELATION

In this section we relate the equal-time correlation

$$\Gamma_{m,\alpha} = \overline{\tilde{c}_1 \cdots \tilde{c}_{2m}}, \qquad (13)$$

of the alternating currents (AC), $\tilde{c}_u$, of $2m$ different detectors of the UHCM setup in Figs. 1 and 2 of the main text, to the moment $\langle : [\hat{n}(\alpha)]^m : \rangle$. In the further calculation the AC $\tilde{c}_u$ is assumed to be obtained by subtracting the mean current $\bar{c}_u$ (DC) from the photocurrent $c_u$. The expression (13) can be related to quantum mechanical expectation values by using photoelectric detection theory,

$$\begin{aligned} \Gamma_{m,\alpha} &= \overline{(c_1 - \bar{c}_1) \cdots (c_{2m} - \bar{c}_{2m})} \\ &= (g_1 \cdots g_{2m})(\eta_1 \cdots \eta_{2m}) \left\langle : [\hat{a}_1^\dagger \hat{a}_1 - \langle \hat{a}_1^\dagger \hat{a}_1 \rangle] \cdots [\hat{a}_{2m}^\dagger \hat{a}_{2m} - \langle \hat{a}_{2m}^\dagger \hat{a}_{2m} \rangle] : \right\rangle. \end{aligned} \qquad (14)$$

Here $\eta_u$ and $g_u$ are the quantum efficiency of the detector $D_u$ and the gain factor of the output current, respectively. Using the relation

$$\hat{a}_u = T_u \hat{a}_0 + \text{vac}, \qquad (15)$$

between the modes $\hat{a}_0$ and $\hat{a}_u$, the latter recorded by detector $D_u$, Eq. (14) can be rewritten as

$$\Gamma_{m,\alpha} = \zeta^{2m} \left\langle : \left[\hat{a}_0^\dagger \hat{a}_0 - \langle \hat{a}_0^\dagger \hat{a}_0 \rangle\right]^{2m} : \right\rangle, \qquad (16)$$

where the balance condition $g_u \eta_u |T_u|^2 = \zeta$, for all $u = 1, \ldots, 2m$, is used. This is easily done by adjusting the $g_u$ values of the amplifiers. The signal $\hat{a}$ is combined with the field of a displaced dephased laser (DDL) at a highly transmitting beam splitter with amplitude transmittance $T$ and reflectance $R$. The DDL is produced by combining

the field $\hat{b}_R$ of a fully dephased coherent state of amplitude $\beta_R \in \mathbb{R}$ with a coherent state of amplitude $\beta_D$ at a beam splitter with amplitude transmittance $T_D$ and reflectance $R_D$. In total this can be described by the relation

$$\hat{a}_0 = T(\hat{a} - \alpha) + RR_D \hat{b}_R, \qquad (17)$$

with $\alpha = -RT_D\beta_D/T$. The corresponding number operator reads as

$$\hat{n}_0 = |T|^2 \hat{n}(\alpha) + |R|^2 |R_D|^2 \hat{b}_R^\dagger \hat{b}_R + T^* R R_D (\hat{a}^\dagger - \alpha^*) \hat{b}_R + T R^* R_D^* (\hat{a} - \alpha) \hat{b}_R^\dagger, \qquad (18)$$

where $\hat{n}(\alpha)$ is the displaced photon number operator $\hat{n}(\alpha) = (\hat{a}^\dagger - \alpha^*)(\hat{a} - \alpha)$. Since the mode $\hat{b}_R$ is in the dephased coherent state, $\int_0^{2\pi} d\phi\, |\beta_R e^{i\phi}\rangle \langle \beta_R e^{i\phi}|/(2\pi)$, one gets for the expectation value of $\hat{n}_0$ the expression

$$\langle \hat{n}_0 \rangle = |T|^2 \langle \hat{n}(\alpha) \rangle + |R|^2 |R_D|^2 \beta_R^2. \qquad (19)$$

Inserting these results into Eq. (16), one may derive

$$\Gamma_{m,\alpha} = \frac{(\zeta|T|)^{2m}}{2\pi} \int_0^{2\pi} d\phi \left\langle : \left[ |T|(\hat{n}(\alpha) - \langle \hat{n}(\alpha) \rangle) + |R||R_D|(\hat{a}^\dagger - \alpha^*)\beta_R e^{i\phi} + |R||R_D|(\hat{a} - \alpha)\beta_R e^{-i\phi} \right]^{2m} : \right\rangle. \qquad (20)$$

In the limit of a strong amplitude $\beta_R$, this expression simplifies to

$$\begin{aligned}\Gamma_{m,\alpha} &= (\zeta|T||R||R_D|\beta_R)^{2m} \frac{1}{2\pi} \int_0^{2\pi} d\phi \left\langle : \left[ (\hat{a}^\dagger - \alpha^*)e^{i\phi} + (\hat{a} - \alpha)e^{-i\phi} \right]^{2m} : \right\rangle \\ &= (\zeta|T||R||R_D|\beta_R)^{2m} \binom{2m}{m} \langle : [\hat{n}(\alpha)]^m : \rangle. \end{aligned} \qquad (21)$$

Introducing the new variable $\tilde{\zeta} = \zeta|T||R||R_D|\beta_R$, one arrives at the final result

$$\Gamma_{m,\alpha} = \binom{2m}{m} \tilde{\zeta}^{2m} \langle : [\hat{n}(\alpha)]^m : \rangle. \qquad (22)$$

## IV. IRRELEVANCE OF UNCORRELATED DARK NOISE IN UHCM

The joint-event statistics, for recording $m_1$ photons at detector $D_1$, $m_2$ photons at detector $D_2$ and so on, with including dark noise $\hat{\nu}_u$ of detector $D_u$, is given by

$$P_{m_1,\ldots,m_M} = \left\langle : \frac{(\eta_1 \hat{n}_1 + \hat{\nu}_1)^{m_1}}{m_1!} e^{-(\eta_1 \hat{n}_1 + \hat{\nu}_1)} \cdots \frac{(\eta_M \hat{n}_M + \hat{\nu}_M)^{m_M}}{m_M!} e^{-(\eta_M \hat{n}_M + \hat{\nu}_M)} : \right\rangle. \qquad (23)$$

In an analogous manner as in Eq. (14) we derive

$$\begin{aligned}\Gamma_{m,\alpha} &= \overline{(c_1 - \bar{c}_1) \cdots (c_{2m} - \bar{c}_{2m})} \\ &= (g_1 \cdots g_{2m}) \left\langle : \prod_{u=1}^{2m} \left[ \eta_u(\hat{n}_u - \langle \hat{n}_u \rangle) + (\hat{\nu}_u - \langle \hat{\nu}_u \rangle) \right] : \right\rangle. \end{aligned} \qquad (24)$$

If the dark counts in different detectors are uncorrelated and obey arbitrary statistics, one obtains

$$\Gamma_{m,\alpha} = (g_1 \cdots g_{2m})(\eta_1 \cdots \eta_{2m}) \langle : (\hat{n}_1 - \langle \hat{n}_1 \rangle) \cdots (\hat{n}_{2m} - \langle \hat{n}_{2m} \rangle) : \rangle. \qquad (25)$$

Obviously, this result is the same as Eq. (14), where the dark noise is ignored. Of course, more care is needed in imperfect settings where dark counts of different detectors may be correlated, e.g., in case of background light or current fluctuations in a common power supply.

## V. ERROR CALCULATION

This section outlines the determination of the error bar for the minimal eigenvalue of the $(k+1) \times (k+1)$ matrix of moments,

$$\left[\underline{\boldsymbol{L}}_w^{(k)}(\alpha)\right]_{mm'} = \mathcal{M}_{w,m+m',\alpha}, \tag{26}$$

where

$$\mathcal{M}_{w,\ell,\alpha} = w^{2\ell}\langle :[\hat{n}(\alpha)]^\ell:\rangle. \tag{27}$$

For this purpose, in a first step the error of the elements of this matrix is calculated in a straightforward manner. The measurement of $\langle :[\hat{n}(\alpha)]^m:\rangle$ requires $2m$ different detectors. Accordingly, to obtain the matrix (26) the number $M$ of detectors has to be greater than or equal to $4k$. As derived in Section III, the equal-time correlation of $2m$ ACs yields the elements of the matrix (26), i.e.,

$$\mathcal{M}_{w,\ell,\alpha} = \frac{\tilde{w}^{2\ell}}{\binom{2\ell}{\ell}}\overline{\tilde{c}_1\cdots\tilde{c}_{2\ell}}. \tag{28}$$

with $\tilde{w} = w/\tilde{\zeta}$. Let $N$ be the number of data samples used to calculate this correlation. Then the sampling error of $\mathcal{M}_{w,\ell,\alpha}$ is calculated as

$$\Delta\mathcal{M}_{w,\ell,\alpha} = \frac{\tilde{w}^{2\ell}}{\sqrt{N}\binom{2\ell}{\ell}}\sqrt{\overline{\tilde{c}_1^2\cdots\tilde{c}_{2\ell}^2} - \left(\overline{\tilde{c}_1\cdots\tilde{c}_{2\ell}}\right)^2}. \tag{29}$$

The first expectation value on the right hand side can be evaluated as

$$\begin{aligned}\overline{\tilde{c}_1^2\cdots\tilde{c}_{2\ell}^2} &= \overline{(c_1-\overline{c}_1)^2\cdots(c_{2\ell}-\overline{c}_{2\ell})^2} \\ &= \sum_{v_1=0}^\infty \cdots \sum_{v_{2\ell}=0}^\infty P_{v_1,\ldots,v_{2\ell}}(g_1[v_1-\overline{v_1}])^2\cdots(g_{2\ell}[v_{2\ell}-\overline{v_{2\ell}}])^2,\end{aligned} \tag{30}$$

where the joint-event statistics,

$$P_{v_1,\ldots,v_{2\ell}} = \left\langle :\frac{(\eta_1\hat{n}_1)^{v_1}}{v_1!}e^{-\eta_1\hat{n}_1}\cdots\frac{(\eta_{2\ell}\hat{n}_{2\ell})^{v_{2\ell}}}{v_{2\ell}!}e^{-\eta_{2\ell}\hat{n}_{2\ell}}:\right\rangle, \tag{31}$$

is the probability to measure $v_1$ photocounts at detector $D_1$, $v_2$ photocounts at detector $D_2$, and so on. It is easy to show by using photoelectric detection theory, the balance condition $g_u\eta_u|T_u|^2 = \zeta$, and relation (15), that (30) simplifies to

$$\begin{aligned}\overline{\tilde{c}_1^2\cdots\tilde{c}_{2\ell}^2} &= (g_1\eta_1)^2\cdots(g_{2\ell}\eta_{2\ell})^2\left\langle:\prod_{u=1}^{2\ell}\left[(\hat{n}_u-\langle\hat{n}_u\rangle)^2 + \hat{n}_u/\eta_u\right]:\right\rangle \\ &= \zeta^{4\ell}\left\langle:\prod_{u=1}^{2\ell}\left[(\hat{n}_0-\langle\hat{n}_0\rangle)^2 + \hat{n}_0/(\eta_u|T_u|^2)\right]:\right\rangle.\end{aligned} \tag{32}$$

For the subsequent calculation, it is advantageous to introduce additionally the symbol

$$\Xi_{\ell,p} = \sum_{\substack{\mathcal{U}\subseteq\{1,\ldots,2\ell\}\\|\mathcal{U}|=2\ell-p}}\prod_{u\in\mathcal{U}}\frac{1}{\eta_u|T_u|^2}, \quad p = 0,\ldots,2\ell. \tag{33}$$

Now, the further evaluation of (32) yields

$$\overline{\tilde{c}_1^2\cdots\tilde{c}_{2\ell}^2} = \zeta^{4\ell}\sum_{p=0}^{2\ell}\Xi_{\ell,p}\left\langle:(\hat{n}_0-\langle\hat{n}_0\rangle)^{2p}\hat{n}_0^{2\ell-p}:\right\rangle. \tag{34}$$

The quantum mechanical expectation values can be related to the signal field by considering the combination (17) of the signal field $\hat{a}$ with the displaced dephased laser at a beam splitter as

$$\begin{aligned}
\overline{\tilde{c}_1^2 \cdots \tilde{c}_{2\ell}^2} &= \zeta^{4\ell} \sum_{p=0}^{2\ell} \Xi_{\ell,p} \frac{1}{2\pi} \int_0^{2\pi} d\phi \\
&\times \Big\langle :\big[|T|^2(\hat{n}(\alpha) - \langle\hat{n}(\alpha)\rangle) + |T||R||R_\mathrm{D}|(\hat{a}^\dagger - \alpha^*)\beta_\mathrm{R} e^{i\phi} + |T||R||R_\mathrm{D}|(\hat{a} - \alpha)\beta_\mathrm{R} e^{-i\phi}\big]^{2p} \\
&\times \big[|T|^2 \hat{n}(\alpha) + |R|^2|R_\mathrm{D}|^2 \beta_\mathrm{R}^2 + |T||R||R_\mathrm{D}|(\hat{a}^\dagger - \alpha^*)\beta_\mathrm{R} e^{i\phi} + |T||R||R_\mathrm{D}|(\hat{a} - \alpha)\beta_\mathrm{R} e^{-i\phi}\big]^{2\ell - p} :\Big\rangle \\
&= \zeta^{4\ell} \sum_{p=0}^{2\ell} \Xi_{\ell,p} \sum_{a_1+a_2+a_3=2p} \binom{2p}{a_1, a_2, a_3} \sum_{b_1+b_2+b_3=2\ell-p} \binom{2\ell - p}{b_1, b_2, b_3} \\
&\times |T|^{2a_1} (-|T|^2 \langle\hat{n}(\alpha)\rangle)^{a_2} (|T||R||R_\mathrm{D}|\beta_\mathrm{R})^{a_3} |T|^{2b_1} (|R|^2|R_\mathrm{D}|^2 \beta_\mathrm{R}^2)^{b_2} (|T||R||R_\mathrm{D}|\beta_\mathrm{R})^{b_3} \\
&\times 1[(a_3+b_3)\text{ even}] \binom{a_3+b_3}{(a_3+b_3)/2} \Big\langle : [\hat{n}(\alpha)]^{a_1+b_1+(a_3+b_3)/2} : \Big\rangle.
\end{aligned} \qquad (35)$$

Here $1[a]$ is 1 if $a$ is true, and 0 otherwise. Inserting Eq. (27) and Eq. (35) into Eq. (29), yields the error $\Delta \mathcal{M}_{w,\ell,\alpha}$. This expression is useful to determine the error without the need for simulating experimental data. For the recorded ACs $\{\tilde{c}_u^{(t)}\}_{t=1}^N$ of $2m$ different detectors $D_u$, each sequence containing $N$ data samples, the sampling error may be determined by the formula

$$\Delta \mathcal{M}_{w,\ell,\alpha} = \sqrt{\frac{1}{N(N-1)} \sum_{t=1}^N \left[ \frac{\tilde{w}^{2\ell}}{\binom{2\ell}{\ell}} \tilde{c}_1^{(t)} \cdots \tilde{c}_{2\ell}^{(t)} - \mathcal{M}_{w,\ell,\alpha} \right]^2}. \qquad (36)$$

Using the error $[\Delta \underline{\boldsymbol{L}}_w^{(k)}(\alpha)]_{mm'} = \Delta \mathcal{M}_{w,m+m',\alpha}$ of the elements of the matrix $\underline{\boldsymbol{L}}_w^{(k)}(\alpha)$, the error of the minimal eigenvalue,

$$\mathcal{F}_w^{(k)}(\alpha) = \min\left\{ \mathrm{S}[\underline{\boldsymbol{L}}_w^{(k)}(\alpha)] \right\}, \qquad (37)$$

of this matrix with the spectrum S, is obtained by performing a linear error propagation. If the unit vector $\boldsymbol{v} \equiv \boldsymbol{v}_w^{(k)}(\alpha)$ is an eigenvector corresponding to the minimal eigenvalue (37), the error of $\mathcal{F}_w^{(k)}(\alpha)$ reads as

$$\Delta \mathcal{F}_w^{(k)}(\alpha) = |\boldsymbol{v}|^T \Delta \underline{\boldsymbol{L}}_w^{(k)}(\alpha) |\boldsymbol{v}|, \qquad (38)$$

where $|\boldsymbol{v}| = (|v_0|, \ldots, |v_k|)$. From Eqs. (37) and (38) one may determine the statistical significance as

$$\Sigma = \begin{cases} |\mathcal{F}_w^{(k)}(\alpha)|/\Delta \mathcal{F}_w^{(k)}(\alpha) & \text{if } \mathcal{F}_w^{(k)}(\alpha) < 0 \\ 0 & \text{else} \end{cases}. \qquad (39)$$

## VI. STATISTICAL SIGNIFICANCE

In the main text a weakly squeezed vacuum state and a single-photon-added thermal state are considered. Figs. 1 and 2 show the statistical significances $\Sigma$ of the negativities of $\mathcal{F}_w^{(k)}(\alpha)$ and $\mathcal{P}_w^{(k)}(\alpha)$ for these states. The significance is determined according to Eq. (39). In Fig. 1 the value of $\Sigma$ is zero for $\mathcal{P}_w^{(2)}(\alpha)$ and in Fig. 2 the only nonvanishing $\Sigma$-value occurs for $\mathcal{F}_w^{(2)}(\alpha)$. A significance of $\Sigma \gtrsim 5$ clearly certifies the nonclassicality of the state.

## VII. AMPLITUDE FLUCTUATIONS OF THE DDL

Since we are dealing with unbalanced homodyne measurements, we must consider amplitude fluctuation effects of the DDL. In this case we replace Eq. (7) in the main part of the Letter with

$$\Gamma_{m,\alpha} = \overline{\tilde{c}_1 \cdots \tilde{c}_{2m}} = \binom{2m}{m} \overline{\tilde{\zeta}^{2m}} \langle : [\hat{n}(\alpha)]^m : \rangle, \qquad (40)$$

$$\overline{\tilde{\zeta}^{2m}} = (\zeta |T||R||R_\mathrm{D}|)^{2m} \overline{\beta_\mathrm{R}^{2m}}. \qquad (41)$$

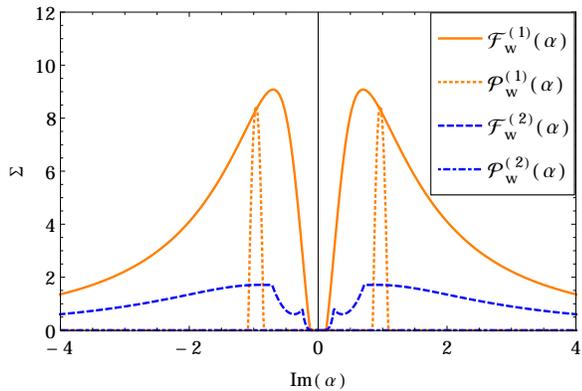
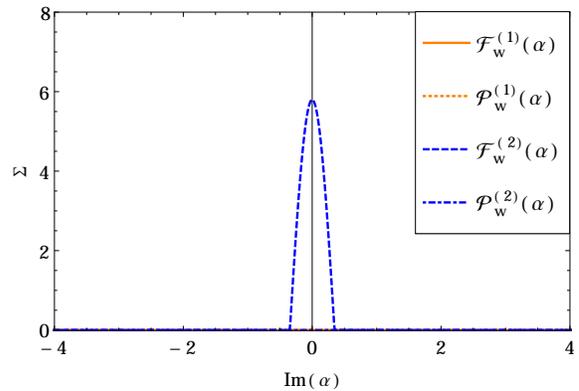

FIG. 1: (Color online) The statistical significance $\Sigma$ for $N = 10^6$ data samples is shown for the minimal eigenvalues $\mathcal{F}_w^{(k)}(\alpha)$ and the truncated regularized $P$ function $\mathcal{P}_w^{(k)}(\alpha)$ for a squeezed vacuum state with $\xi = 0.03$ along the squeezed axis, for $q = 10$ and $w = 1.5$. The functions are illustrated for $k = 1$ and $k = 2$.

FIG. 2: (Color online) The statistical significance $\Sigma$ for $N = 10^7$ data samples is shown for the minimal eigenvalues $\mathcal{F}_w^{(k)}(\alpha)$ and the truncated regularized $P$ function $\mathcal{P}_w^{(k)}(\alpha)$ for the SPATS with the parameters described in the main text along the imaginary axis, for $q = 10$ and $w = 1.3$. The functions are illustrated for $k = 1$ and $k = 2$.

Let us assume that the amplitude fluctuations, $\Delta\beta_\mathrm{R} = \beta_\mathrm{R} - \overline{\beta_\mathrm{R}}$, are Gaussian. In typical experiments, the relative intensity fluctuations can be stabilized to about $10^{-2}$, over time intervals of the order needed for state reconstruction. For moments of increasing order the effects of amplitude noise are increasing. The examples in the Letter require moments of fourth order, for which we consider the noise effects. For Gaussian noise the leading terms are

$$\overline{\beta_\mathrm{R}^4} \approx \overline{\beta_\mathrm{R}}^4 \left(1 + 28 \frac{\overline{\Delta\beta_\mathrm{R}^2}}{\overline{\beta_\mathrm{R}}^2}\right) \approx \overline{\beta_\mathrm{R}}^4 (1 + 7 \times 10^{-4}). \tag{42}$$

Hence for moments up to fourth order it is still a very good approximation to ignore the amplitude fluctuations. In case of higher-order moments the amplitude fluctuations can be recorded in an auxiliary channel in Fig. 1 of the Letter, and the measured correlations according to Eq. (40) can be corrected by dividing with the recorded $\overline{\beta_\mathrm{R}^{2m}}$ values. The relative intensity fluctuations of the DDL are further reduced by correcting the data within sufficiently small time intervals.

---


\end{widetext}

\end{document}